\documentclass[sigconf,natbib=true]{acmart}
\setcopyright{rightsretained}
\acmConference[SIGIR eCom'22]{ACM SIGIR Workshop on eCommerce}{July 15, 2022}{ Madrid, Spain}
\acmYear{2022}
\copyrightyear{2022}
\makeatletter
\renewcommand\@formatdoi[1]{\ignorespaces}
\makeatother
\acmISBN{}
\AtBeginDocument{%
  \providecommand\BibTeX{{%
    \normalfont B\kern-0.5em{\scshape i\kern-0.25em b}\kern-0.8em\TeX}}}

\setcopyright{acmcopyright}
\copyrightyear{2022}
\acmYear{2022}

\acmConference[SIGIR eCom'22]{SIGIR eCom'22, SIGIR Workshop on eCommerce}{July 15 2022}{Madrid, Spain}

\usepackage{todonotes}
\usepackage{adjustbox}
\usepackage{amsmath}
\usepackage{xcolor}

\begin{document}

\title{Incorporating Customer Reviews in Size and Fit Recommendation systems for Fashion E-Commerce}

\author{Oishik Chatterjee}
\affiliation{%
  \institution{Flipkart Internet Private Limited}
  \city{Bengaluru}
  \country{India}}
  \email{oishik.chatterjee@flipkart.com}

\author{Jaidam Ram Tej}
\affiliation{%
  \institution{Flipkart Internet Private Limited}
  \city{Bengaluru}
  \country{India}}
\email{jaidam.ramtej@flipkart.com}

\author{Narendra Varma Dasaraju}
\affiliation{%
 \institution{Flipkart Internet Private Limited}
 \city{Bengaluru}
 \country{India}}
 \email{narendra.varma@flipkart.com}

\renewcommand{\shortauthors}{Oishik, Jaidam and Narendra, et al.}

\begin{abstract}

With the huge growth in e-commerce domain, product recommendations have become an increasing field of interest amongst e-commerce companies. One of the more difficult tasks in product recommendations is size and fit predictions. There are a lot of size related returns and refunds in e-fashion domain which causes inconvenience to the customers as well as costs the company. Thus having a good size and fit recommendation system, which can predict the correct sizes for the customers will not only reduce size related returns and refunds but also improve customer experience. Early works in this field used traditional machine learning approaches to estimate customer and product sizes from purchase history. These methods suffered from cold start problem due to huge sparsity in the customer-product data. More recently, people have used deep learning to address this problem by embedding customer and product features. But none of them incorporates valuable customer feedback present on product pages along with the customer and product features. We propose a novel approach which can use information from customer reviews along with customer and product features for size and fit predictions. We demonstrate the effectiveness of our approach compared to using just product and customer features on 4 datasets. Our method shows an improvement of 1.37\% - 4.31\%  in F1 (macro) score over the baseline across the 4 different datasets.

\end{abstract}

\maketitle

\section{Introduction}
Online fashion market is expected to grow at 11.4\% per year \cite{ecommercereport2020}. Returns, where size issue is a considerable piece of the pie, are the bane for the fashion e-commerce companies. Up to 40\% of online fashion products are returned \cite{dogani2019learning}. In in-store purchases, consumers prefer to see, touch, and try-on apparel before purchasing. They lack similar engaging experience in online shopping.

In online shopping, consumers rely on symbolic sizes (e.g. ‘S’, ‘M’, ‘L’) to make their purchase decisions. Though symbolic sizes are mentioned in products, they vary between brands. Sometimes there is size variation within a brand \cite{sheikh2019deep}. Consumers also use size guides. Size guides provide a mapping from the standard sizes to corresponding physical sizes, in cm or inches. There are multiple symbolic sizing schemes (e.g. ‘US’, ‘UK’, ‘EU’) in a size guide. These size guides are usually at a brand level and do not capture finer fit details of a product. They are cumbersome to enact and require
measuring instruments at their disposal. Further, due to vanity sizing, consuming symbolic sizes can be tricky. Thus, sizes mentioned on the products are no longer enough to make a purchase.

Customers when buying a product look for the products with the right fit. They usually return the product if it does not fit. It is important that E-commerce platforms provide accurate recommendation tips to customers. This helps the customers in three ways. Firstly it helps in reducing the customers' returns. Secondly, it eases customers in finding their right fit enriching their shopping experience online and hence might boost conversions. Thirdly it helps in building customer loyalty.  Hence we need a good size and fit recommendation system which can help customers in choosing the right fit.

There have been various solutions proposed for size and fit recommendation. Most of the earlier solutions used traditional approaches to embed customer product transactions to determine the right size for the customers. These approaches suffer from the cold start problem because the customer-product transaction data is sparse. Recently deep learning approaches \cite{sheikh2019deep,dogani2019learning, eshel2021presize} which use customer and product features along with transaction data have been proposed which tries to mitigate this problem. None of the current approaches use the customer reviews present on the product page. These reviews often contain information that can help in predicting size/fit of new customers. When customers return a product they might give some information in the review which indicates the size/fit of the product. Though \cite{DBLP:journals/corr/abs-1908-10896} uses reviews to do fit prediction, they only classify each review as small fit or large. Such a model alone cannot help in recommending size for a new customer. Hence we propose a Deep Learning based approach that uses customer reviews along with product and customer information to predict the right fit.

\textbf{Our contributions are} :
\begin{itemize}
    \item We propose a novel approach of leveraging size and fit information in customer reviews present on E-Commerce platforms for size and fit predictions.
    \item We demonstrate how user reviews given by customer on product pages of E-Commerce platforms can be embedded using state-of-the-art pre-trained language models (such as BERT \cite{devlin2018bert}) and used along with product and customer features for improving size and fit prediction models. To our knowledge, this is the first work which utilizes both product-customer features along with customer reviews for doing size and fit predictions.
    \item We empirically show on 4 datasets curated using the data of one of the largest e-commerce platforms that using size and fit information present in customer reviews does indeed help in predicting correct fit for new customers.
\end{itemize}

The rest of the paper is organized as follows. Section 2 describes the related work. Section 3 describes the problem formulation for size and fit recommendation system. Section 4 describes our experiments and datasets used. Experimental results are reported in Section 5 and finally we conclude in section 6.

\section{Related work}
In literature, of late there has been a lot of focus on the size and fit problem \cite{sembium2017recommending,sembium2018bayesian}. Abdulla et al. \cite{mohammed2019shop} embed both users and products using skip-gram based Word2Vec model \cite{mikolov2013efficient} and employ GBM classifier \cite{friedman2001greedy} to predict the fit. A latent factor model was proposed by Sembium et al. \cite{sembium2017recommending}, which was later follow-up by a Bayesian formulation \cite{sembium2018bayesian} to predict the size of a product (small, fit, large). In \cite{sembium2018bayesian} Bayesian logistic regression with ordinal categories was used. They proposed an efficient algorithm for posterior inference based on mean-field variational inference and Polya-Gamma augmentation. Guigourès et al. employed a hierarchical Bayesian model \cite{guigoures2018hierarchical} for personalized size recommendation. Misra et.al. \cite{misra2018decomposing} learn the fit semantics by modeling it as an ordinal regression problem. Then, they employ metric learning techniques to address the class imbalance issues. 

Recently, deep learning approaches have been used to solve the size recommendation problem with encouraging results \cite{lasserre2020meta,sheikh2019deep,dogani2019learning}. Deep Learning approaches unlike the traditional approaches are able to scale well with large amounts of data. SFNET \cite{sheikh2019deep} provides recommendations at the user cross product level using a deep learning based content collaborative approach. The approach can learn from cross-correlations that exist across fashion categories. They use both purchase and returns data as well as customer and article features for personalized size and fit prediction. Dogani et al. \cite{dogani2019learning} addressed the sparsity problem by learning latent representation at a brand level using neural collaborative filtering \cite{he2017neural}. Then, fine-tuning the product representation by transfer learning from brand representation. Lasserre et al. \cite{lasserre2020meta} use a deep learning based meta learning approach. Their approach is based on the premise that, given the purchase history of a customer \textit{i}, products \textit{\(x_j\)} and their corresponding size estimates \textit{\(y_{ij})\)} share a strong linear relationship. Baier et al. \cite{DBLP:journals/corr/abs-1908-10896} derive product fit feedback from customer reviews using natural language processing techniques which is then used to infer the right fit.


A few approaches explore the use of product images or 3D scan of products to predict the right fit for a customer. SizeNet \cite{karessli2019sizenet} uses product images to infer whether the product will have fit issues for a customer. 
ViBE uses a computer vision approach to develop a body-aware embedding that captures garment’s affinity with different body shapes \cite{hsiao2020vibe}. The approach learns the embedding from images of models of various shapes and sizes wearing the product, displayed on catalog. \cite{eshel2021presize} proposes PreSizE – a  size prediction framework that utilizes Transformers to capture the relationship between various item attributes (e.g., brand, category, etc.) and its purchased size by encoding items and user’s purchase history. 

In industrial applications, customers are looped in for size recommendation. These applications ask targeted questions to acquire biometric measurements, e.g. height, weight, age, waist of the customer. They also ask queries related to previous purchases, e.g. “Which brand and size of shoe do you find most comfortable?”. These data points are further used to predict for cold start customers. However, the downside is that it may add friction in the consumer path.

In this work, we focus on using customer reviews, purchases and returns data of the customers, product information in predicting the right fit for a customer. Customer reviews might contain crucial size and fit information which might help in predicting the right fit to a customer over using product and customer information alone. To extract the information from customer reviews we have embedded the reviews using a pre-trained language model \cite{devlin2018bert}. A more detailed description of the model is discussed in Section 3. 


\begin{figure}[h]
  \centering
  \includegraphics[width=.75\linewidth]{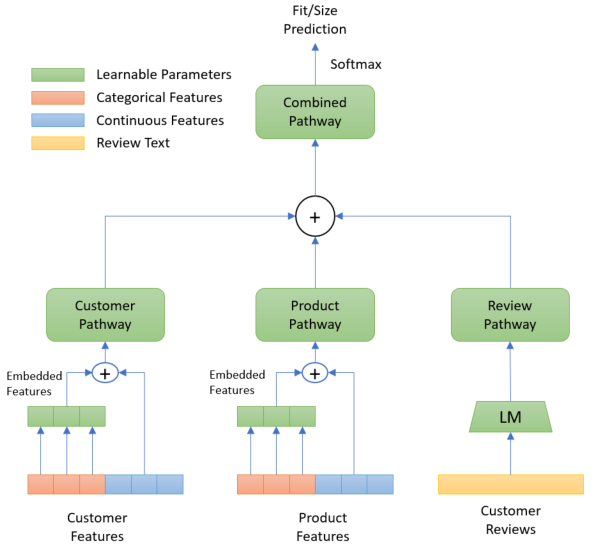}
  \caption{Architecture}
  \Description{Architecture}
  \label{fig : deep_neural_net}
  \vspace{-5mm}
\end{figure}

\begin{table*}[h]
  \caption{Dataset Statistics}
  \label{tab:dataset statistics}
  \begin{tabular}{c|ccccl}
    \toprule
    \textbf{Description}&\textbf{Women Jacket}&\textbf{Women Kurta and Kurti}&\textbf{Mens Jean}&\textbf{Mens Polo Tshirt} \\
    \midrule
    \midrule
    \textbf{No. of Customers} &14,94,713  &61,53,598 &31,72,637  &21,18,089\\
    \textbf{No. of Products} &47,040  &1,20,425  &47,140  &47,040\\
    \textbf{\% "Small" Instances} &4.52\% &2.40\% &2.92\% &3.14\%\\
    \textbf{\% "Large" Instances} &1.68\% &1.30\% &2.35\% &2.76\%\\
    
  \bottomrule
\end{tabular}
\end{table*}

\section{Size and Fit Recommendation}

We model the size and fit problem as a classification problem where given a product and a customer we want to predict if the product will 'fit' the customer, or it will be 'small' or 'large' for the customer. The following subsections explain the problem formulation and the model architecture.

\subsection{Problem Formulation}
\label{subsection : problem_formn}

 Given a customer $\textbf{C}$, a product $\textbf{P}$, and a set of reviews for the product $\textbf{R}$, we want to predict if the product will be small, fit or large for the customer. Both product and customer are defined by their respective features $\textbf{P} = \{p_i\}$ and $\textbf{C} = \{c_i\}$ where each feature can be either continuous or categorical. $\textbf{R} = \{r_i\}$ consists of the reviews left by customers on the product page. We define the output space as $\textbf{F} = \{small,\  fit,\  large\}$. We want to learn the probability distribution of $\textbf{F}$ given $\textbf{C}$, $\textbf{P}$, and $\textbf{R}$ i.e. $p_\theta(f|\textbf{C}, \textbf{P},\textbf{R})$ where $\theta$ denotes the model parameters.
 
\subsection{Neural Network architecture}

The network architecture consists of 3 input pathways for customer, product and review inputs which are then followed by a combined pathway that outputs the final prediction. Each pathway consists of a series of residual blocks \cite{he2016deep}. 

Firstly for the product input features, the numerical features like size are normalized and the categorical features like brand, fabric etc are converted to vectors using embedding layers. These are then concatenated together ($\boldsymbol{h_p}$) and passed through a residual block (product input pathway) to generate product embedding ($\boldsymbol{e_p}$). The same is done for the customer features to generate customer embedding($\boldsymbol{e_c})$. For the reviews, each review is encoded using a pre-trained language model (more details in section \ref{subsection : review_embeddings}). The embeddings are then averaged and passed through another residual block \cite{he2016deep} similar to the ones used for customer and product input pathways.

\begin{equation}
    h_{pi} =
    \begin{cases}
      NormalizationLayer(p_i) & \text{if $p_i$ is numerical}\\
      EmbeddingLayer(p_i) & \text{if $p_i$ is categorical}
    \end{cases}
\end{equation}







\begin{align}
    \boldsymbol{h_p} &= \bigoplus h_{pi} \\
    \boldsymbol{e_p} &= \text{Product Input Pathway}(\boldsymbol{h_p}) \\
    h_{ri} &= \text{LanguageModel}(r_i) \\
    \boldsymbol{h_r} &= \text{Mean}(h_{ri}) \\
    \boldsymbol{e_r} &= \text{Review Input Pathway}(\boldsymbol{h_r})
\end{align}

\begin{equation}
    h_{ci} =
    \begin{cases}
      NormalizationLayer(c_i) & \text{if $c_i$ is numerical}\\
      EmbeddingLayer(c_i) & \text{if $c_i$ is categorical}
    \end{cases}
\end{equation}


\begin{align}
    \boldsymbol{h_c} &= \bigoplus h_{ci} \\
    \boldsymbol{e_c} &= \text{Customer Input Pathway}(\boldsymbol{h_c})
\end{align}

The embedding from the pathways are combined as \([\boldsymbol{e_c}, \boldsymbol{e_p}, \boldsymbol{e_r} ]\) and passed through the combined pathway. The combined pathway again consists of a series of residual blocks which is followed by a softmax layer. 

\begin{align}
    \boldsymbol{e} &=\boldsymbol{e_c} \oplus \boldsymbol{e_p} \oplus \boldsymbol{e_r}\\
    \boldsymbol{o} &=\text{Combined Pathway}(\boldsymbol{e})\\
    p_\theta(f|\textbf{C}, \textbf{P},\textbf{R}) &= \text{softmax} (\boldsymbol{o}) \ \  f\in F\\
    y &=  \mathop{argmax}  \limits_{f\in F} \  p_\theta(f|\textbf{C}, \textbf{P},\textbf{R})
\end{align}


The number of output labels is related to the reason codes provided by the customer. Customers return products when there are size and fit issues. They select the following reason codes: ‘Size smaller’ or ‘Size larger’. Based on this premise our size recommendations are of the form ‘buy one size small’ or ‘buy one size large’. When the product is true to size no recommendation is shared with the customer. In total there are three classes we wish to determine; small, large and fit corresponding to ‘buy one size large’, ‘buy one size small’ and true to size, respectively.

\begin{table}
  \caption{Model Learning Parameters}
  \label{tab:hyperparameters}
  \begin{tabular}{ccl}
    \toprule
    \textbf{Parameter}&\textbf{Details}\\
    \midrule
    \midrule
    Batchsize & 2048 \\
    Learning rate & 0.01\\
    Optimizer & Adam\\
    Review embedding dimension & 768\\
    Input Pathways & \#(emb + cont) feat x 25 x 15 x 10 \\
    Final Pathway &  50 x 100 x 200 x 500 x 3 \\
  \bottomrule
\end{tabular}
\end{table}

\begin{table*}[h]
  \caption{F1-Macro score of all the models (OR: Only Reviews, BM: BERT-Modcloth, BFC: BERT-FC, FB: FKBERT, FFC: FKBERT-FC). The numbers denote the improvement in \% over the baseline(SFNet).}
  \label{table:results}
  \begin{tabular}{c|cccl}
    \toprule
    \textbf{Model}&\textbf{Women Jacket}& \textbf{Women Kurta and Kurti}& \textbf{Mens Jean}& \textbf{Mens Polo Tshirt}\\
    \midrule
    \midrule
    \textbf{SFNet (Baseline)} &38.26  &35.6  &37.19  & \  37.2 \\
    \midrule
    \textbf{OR} & -15.68\% &-8.13\% &-12.77\% &-13.12\%\\
    \textbf{BM} &+0.68\% &+1.94\% &+0.83\% &+0.43\%\\
    \textbf{BFC}  &+3.27\% &+2.81\% &+1.00\% &+2.15\%\\
    \textbf{FB} &+1.93\% &+1.57\% &\textbf{+1.37}\% &+2.34\%\\
    \textbf{FFC} &\textbf{+4.31}\% &\textbf{+3.09}\% &\textbf{+1.37}\% &\textbf{+2.42}\%\\
    
  \bottomrule
\end{tabular}
\end{table*}

\subsection{Review Embeddings}
\label{subsection : review_embeddings}

We embed reviews using a pre-trained language model before passing on to  review input pathway. We use BERT \cite{devlin2018bert} as the pre-trained language model here (we have experimented with different language models and found that BERT \cite{devlin2018bert} gives the best results). We pre-train BERT \cite{devlin2018bert} in a variety of ways and report numbers on each pre-trained model. We train some of the BERT \cite{devlin2018bert} model on review fit classification task where each review is labeled \{small, fit, large\} which helps the model learn about the fit information present in the review text that can help in subsequent size and fit prediction task. More details is given in section \ref{subsection : models}. Each product may have zero, one or many reviews. Since we want a single review feature for each product, therefore in case of many reviews we aggregate the review embedding vectors of all the reviews for a given product to get a single vector. To do this we average over all the embedding vectors. In case of zero reviews we use a default vector of all zeros.


\section{Experiments}
\label{section : experiments}
We demonstrate that size and fit prediction models can leverage information present in customer reviews to improve their performance by comparing our approach with SFNet \cite{sheikh2019deep} which uses only customer and product features on 4 different datasets (each a separate category) - \textbf{Women Jacket}, \textbf{Women Kurta and Kurti}, \textbf{Mens Jean} and \textbf{Mens Polo Tshirt}.

\subsection{Dataset}
\label{subsection : dataset}
We have collected product details, purchase and returns data, and customer reviews for the following categories: \textbf{Women Jacket}, \textbf{Women Kurta and Kurti}, \textbf{Mens Jean} and \textbf{Mens Polo Tshirt}. We label each purchase with small or large based on if the customer has returned the product stating small size or large size as the reason respectively. If the customer has not returned the product, then we label it as fit. Table \ref{tab:dataset statistics} shows the data statistics of all the categories we have trained on. We have split each dataset in 80:10:10 train validation and test split.
We also create a reviews dataset for each category which is used to train the BERT model \cite{devlin2018bert} for embedding the customer reviews. Here each review is labeled as small, fit or large based on the criteria mentioned above. We also use a public dataset - Modcloth \cite{misra2018decomposing}, for training the BERT model \cite{devlin2018bert} to show the effectiveness of transfer learning when labeled reviews dataset is not available. 

\subsection{Models}
\label{subsection : models}
We have trained the language model in the following ways:
\begin{itemize}
    \item \textbf{BERT-Mocloth}: We have trained a BERT model \cite{devlin2018bert} on the fit classification task with the Modcloth reviews dataset.
    \item \textbf{BERT-FC}: We have trained a BERT model \cite{devlin2018bert} on the fit classification task with the reviews of the given dataset.
    \item \textbf{FKBERT}: We have trained a ROBERTa model \cite{liu2019roberta} on various text e-commerce data such as product description, reviews, question-answers, etc. on the masked language modeling task. 
    \item \textbf{FKBERT-FC}: We have fine-tuned the FKBERT model on the fit classification task with reviews of the given dataset.
\end{itemize}
We also report numbers on \textbf{Only Reviews} which uses the predictions by the bert model directly and determines the class of the product using majority voting of the review predictions. For the baseline we compare against \textbf{SFNet} model \cite{sheikh2019deep} which uses just customer and product features for prediction.

\subsection{Hyperparameters, Training and Evaluation}
We have used the same hyperparameters for all the datasets. Details of the hyperparameters are given in Table \ref{tab:hyperparameters}. 

We have first trained the BERT \cite{devlin2018bert} model on the reviews dataset. We have used AdamW \cite{loshchilov2017decoupled} optimizer with a learning rate set to 2e-5 for training the BERT model. 

For training the rest of the model, we have used Adam \cite{kingma2014adam} optimizer with learning rate 1e-2. Since the dataset is highly skewed towards the fit class, hence we have re-sampled the fit class instances such that the number of fit instances in the training data is equal to the number of small + large instances. This prevents the model from over-fitting to the fit class. The model is trained for a maximum of 100 epochs and the epoch with the best validation F1 is chosen. For evaluation, we report the F1-macro scores. 

\section{Results}
\label{section : experimantal_results}
We perform a set of experiments on the 4 different datasets mentioned in section \ref{subsection : dataset}. Our baseline model is the SFNET \cite{sheikh2019deep} model which does not use customer reviews. We report the improvement of F1 (Macro) score over the baseline (SFNet \cite{sheikh2019deep}) on the 4 models discussed in \ref{subsection : models} in table \ref{table:results}. Results show that there is a significant improvement when reviews are used with other features. Even without labeled reviews data, we get \textbf{0.43\% - 1.94\%} improvement on the baseline model using transfer learning from public datasets. Also, we observe that pre-training with e-commerce textual data also improves performance as can be seen from \textbf{FKBERT} and \textbf{FKBERT-FC} numbers. The biggest improvement is shown when the language model is first pre-trained on generic e-commerce data and then fine-tuned on the fit classification task. \textbf{FKBERT-FC} shows \textbf{1.37\% - 4.31\%} improvement in macro F1 over the baseline. We also observe that reviews by themselves cannot be used alone for fit prediction as shown by the low numbers of the \textbf{Only Reviews} model.

\section{Conclusion and Future Work}
Size and Fit recommendation is an important problem in e-commerce because it helps customers in choosing the right fit and thereby reduces size and fit related returns. Most of the earlier works embedded user and product information to predict the right fit. In this paper, we propose a deep learning based approach to predict the fit based on customer reviews along with customer and  product information. Through extensive experimentation on different datasets curated from data of one of the largest e-commerce platforms, we show the effectiveness of our approach. 


We plan to extend this work by including other customer feedback like customer uploaded images and customer QnA to predict the right fit. Customers sometimes ask questions related to size and fit which get answered by other customers who have already bought the product. Also using customer uploaded images along with product information might help the model to learn the interactions between user and product to predict the right fit.

\bibliographystyle{ACM-Reference-Format}
\bibliography{sample-base}


\begin{thebibliography}{21}


\ifx \showCODEN    \undefined \def \showCODEN     #1{\unskip}     \fi
\ifx \showDOI      \undefined \def \showDOI       #1{#1}\fi
\ifx \showISBNx    \undefined \def \showISBNx     #1{\unskip}     \fi
\ifx \showISBNxiii \undefined \def \showISBNxiii  #1{\unskip}     \fi
\ifx \showISSN     \undefined \def \showISSN      #1{\unskip}     \fi
\ifx \showLCCN     \undefined \def \showLCCN      #1{\unskip}     \fi
\ifx \shownote     \undefined \def \shownote      #1{#1}          \fi
\ifx \showarticletitle \undefined \def \showarticletitle #1{#1}   \fi
\ifx \showURL      \undefined \def \showURL       {\relax}        \fi
\providecommand\bibfield[2]{#2}
\providecommand\bibinfo[2]{#2}
\providecommand\natexlab[1]{#1}
\providecommand\showeprint[2][]{arXiv:#2}

\bibitem[\protect\citeauthoryear{Baier}{Baier}{2019}]%
        {DBLP:journals/corr/abs-1908-10896}
\bibfield{author}{\bibinfo{person}{Stephan Baier}.}
  \bibinfo{year}{2019}\natexlab{}.
\newblock \showarticletitle{Analyzing Customer Feedback for Product Fit
  Prediction}.
\newblock \bibinfo{journal}{\emph{CoRR}}  \bibinfo{volume}{abs/1908.10896}
  (\bibinfo{year}{2019}).
\newblock
\showeprint[arXiv]{1908.10896}
\urldef\tempurl%
\url{http://arxiv.org/abs/1908.10896}
\showURL{%
\tempurl}


\bibitem[\protect\citeauthoryear{Devlin, Chang, Lee, and Toutanova}{Devlin
  et~al\mbox{.}}{2018}]%
        {devlin2018bert}
\bibfield{author}{\bibinfo{person}{Jacob Devlin}, \bibinfo{person}{Ming-Wei
  Chang}, \bibinfo{person}{Kenton Lee}, {and} \bibinfo{person}{Kristina
  Toutanova}.} \bibinfo{year}{2018}\natexlab{}.
\newblock \showarticletitle{Bert: Pre-training of deep bidirectional
  transformers for language understanding}.
\newblock \bibinfo{journal}{\emph{arXiv preprint arXiv:1810.04805}}
  (\bibinfo{year}{2018}).
\newblock


\bibitem[\protect\citeauthoryear{Dogani, Tomassetti, Vargas, Chamberlain, and
  De~Cnudde}{Dogani et~al\mbox{.}}{2019}]%
        {dogani2019learning}
\bibfield{author}{\bibinfo{person}{Kallirroi Dogani}, \bibinfo{person}{Matteo
  Tomassetti}, \bibinfo{person}{Sa{\'u}l Vargas},
  \bibinfo{person}{Benjamin~Paul Chamberlain}, {and} \bibinfo{person}{Sofie
  De~Cnudde}.} \bibinfo{year}{2019}\natexlab{}.
\newblock \showarticletitle{Learning Embeddings for Product Size
  Recommendations.}. In \bibinfo{booktitle}{\emph{eCOM@ SIGIR}}.
\newblock


\bibitem[\protect\citeauthoryear{Eshel, Levi, Roitman, and Nus}{Eshel
  et~al\mbox{.}}{2021}]%
        {eshel2021presize}
\bibfield{author}{\bibinfo{person}{Yotam Eshel}, \bibinfo{person}{Or Levi},
  \bibinfo{person}{Haggai Roitman}, {and} \bibinfo{person}{Alexander Nus}.}
  \bibinfo{year}{2021}\natexlab{}.
\newblock \showarticletitle{PreSizE: Predicting Size in E-Commerce using
  Transformers}.
\newblock \bibinfo{journal}{\emph{arXiv preprint arXiv:2105.01564}}
  (\bibinfo{year}{2021}).
\newblock


\bibitem[\protect\citeauthoryear{Friedman}{Friedman}{2001}]%
        {friedman2001greedy}
\bibfield{author}{\bibinfo{person}{Jerome~H Friedman}.}
  \bibinfo{year}{2001}\natexlab{}.
\newblock \showarticletitle{Greedy function approximation: a gradient boosting
  machine}.
\newblock \bibinfo{journal}{\emph{Annals of statistics}}
  (\bibinfo{year}{2001}), \bibinfo{pages}{1189--1232}.
\newblock


\bibitem[\protect\citeauthoryear{Guigour{\`e}s, Ho, Koriagin, Sheikh, Bergmann,
  and Shirvany}{Guigour{\`e}s et~al\mbox{.}}{2018}]%
        {guigoures2018hierarchical}
\bibfield{author}{\bibinfo{person}{Romain Guigour{\`e}s},
  \bibinfo{person}{Yuen~King Ho}, \bibinfo{person}{Evgenii Koriagin},
  \bibinfo{person}{Abdul-Saboor Sheikh}, \bibinfo{person}{Urs Bergmann}, {and}
  \bibinfo{person}{Reza Shirvany}.} \bibinfo{year}{2018}\natexlab{}.
\newblock \showarticletitle{A hierarchical bayesian model for size
  recommendation in fashion}. In \bibinfo{booktitle}{\emph{Proceedings of the
  12th ACM Conference on Recommender Systems}}. \bibinfo{pages}{392--396}.
\newblock


\bibitem[\protect\citeauthoryear{He, Zhang, Ren, and Sun}{He
  et~al\mbox{.}}{2016}]%
        {he2016deep}
\bibfield{author}{\bibinfo{person}{Kaiming He}, \bibinfo{person}{Xiangyu
  Zhang}, \bibinfo{person}{Shaoqing Ren}, {and} \bibinfo{person}{Jian Sun}.}
  \bibinfo{year}{2016}\natexlab{}.
\newblock \showarticletitle{Deep residual learning for image recognition}. In
  \bibinfo{booktitle}{\emph{Proceedings of the IEEE conference on computer
  vision and pattern recognition}}. \bibinfo{pages}{770--778}.
\newblock


\bibitem[\protect\citeauthoryear{He, Liao, Zhang, Nie, Hu, and Chua}{He
  et~al\mbox{.}}{2017}]%
        {he2017neural}
\bibfield{author}{\bibinfo{person}{Xiangnan He}, \bibinfo{person}{Lizi Liao},
  \bibinfo{person}{Hanwang Zhang}, \bibinfo{person}{Liqiang Nie},
  \bibinfo{person}{Xia Hu}, {and} \bibinfo{person}{Tat-Seng Chua}.}
  \bibinfo{year}{2017}\natexlab{}.
\newblock \showarticletitle{Neural collaborative filtering}. In
  \bibinfo{booktitle}{\emph{Proceedings of the 26th international conference on
  world wide web}}. \bibinfo{pages}{173--182}.
\newblock


\bibitem[\protect\citeauthoryear{Hsiao and Grauman}{Hsiao and Grauman}{2020}]%
        {hsiao2020vibe}
\bibfield{author}{\bibinfo{person}{Wei-Lin Hsiao} {and}
  \bibinfo{person}{Kristen Grauman}.} \bibinfo{year}{2020}\natexlab{}.
\newblock \showarticletitle{ViBE: Dressing for diverse body shapes}. In
  \bibinfo{booktitle}{\emph{Proceedings of the IEEE/CVF Conference on Computer
  Vision and Pattern Recognition}}. \bibinfo{pages}{11059--11069}.
\newblock


\bibitem[\protect\citeauthoryear{Karessli, Guigour{\`e}s, and
  Shirvany}{Karessli et~al\mbox{.}}{2019}]%
        {karessli2019sizenet}
\bibfield{author}{\bibinfo{person}{Nour Karessli}, \bibinfo{person}{Romain
  Guigour{\`e}s}, {and} \bibinfo{person}{Reza Shirvany}.}
  \bibinfo{year}{2019}\natexlab{}.
\newblock \showarticletitle{Sizenet: Weakly supervised learning of visual size
  and fit in fashion images}. In \bibinfo{booktitle}{\emph{Proceedings of the
  IEEE/CVF Conference on Computer Vision and Pattern Recognition Workshops}}.
  \bibinfo{pages}{0--0}.
\newblock


\bibitem[\protect\citeauthoryear{Kingma and Ba}{Kingma and Ba}{2014}]%
        {kingma2014adam}
\bibfield{author}{\bibinfo{person}{Diederik~P Kingma} {and}
  \bibinfo{person}{Jimmy Ba}.} \bibinfo{year}{2014}\natexlab{}.
\newblock \showarticletitle{Adam: A method for stochastic optimization}.
\newblock \bibinfo{journal}{\emph{arXiv preprint arXiv:1412.6980}}
  (\bibinfo{year}{2014}).
\newblock


\bibitem[\protect\citeauthoryear{Lasserre, Sheikh, Koriagin, Bergman, Vollgraf,
  and Shirvany}{Lasserre et~al\mbox{.}}{2020}]%
        {lasserre2020meta}
\bibfield{author}{\bibinfo{person}{Julia Lasserre},
  \bibinfo{person}{Abdul-Saboor Sheikh}, \bibinfo{person}{Evgenii Koriagin},
  \bibinfo{person}{Urs Bergman}, \bibinfo{person}{Roland Vollgraf}, {and}
  \bibinfo{person}{Reza Shirvany}.} \bibinfo{year}{2020}\natexlab{}.
\newblock \showarticletitle{Meta-learning for size and fit recommendation in
  fashion}. In \bibinfo{booktitle}{\emph{Proceedings of the 2020 SIAM
  international conference on data mining}}. SIAM, \bibinfo{pages}{55--63}.
\newblock


\bibitem[\protect\citeauthoryear{Liu, Ott, Goyal, Du, Joshi, Chen, Levy, Lewis,
  Zettlemoyer, and Stoyanov}{Liu et~al\mbox{.}}{2019}]%
        {liu2019roberta}
\bibfield{author}{\bibinfo{person}{Yinhan Liu}, \bibinfo{person}{Myle Ott},
  \bibinfo{person}{Naman Goyal}, \bibinfo{person}{Jingfei Du},
  \bibinfo{person}{Mandar Joshi}, \bibinfo{person}{Danqi Chen},
  \bibinfo{person}{Omer Levy}, \bibinfo{person}{Mike Lewis},
  \bibinfo{person}{Luke Zettlemoyer}, {and} \bibinfo{person}{Veselin
  Stoyanov}.} \bibinfo{year}{2019}\natexlab{}.
\newblock \showarticletitle{Roberta: A robustly optimized bert pretraining
  approach}.
\newblock \bibinfo{journal}{\emph{arXiv preprint arXiv:1907.11692}}
  (\bibinfo{year}{2019}).
\newblock


\bibitem[\protect\citeauthoryear{Loshchilov and Hutter}{Loshchilov and
  Hutter}{2017}]%
        {loshchilov2017decoupled}
\bibfield{author}{\bibinfo{person}{Ilya Loshchilov} {and}
  \bibinfo{person}{Frank Hutter}.} \bibinfo{year}{2017}\natexlab{}.
\newblock \showarticletitle{Decoupled weight decay regularization}.
\newblock \bibinfo{journal}{\emph{arXiv preprint arXiv:1711.05101}}
  (\bibinfo{year}{2017}).
\newblock


\bibitem[\protect\citeauthoryear{Mikolov, Chen, Corrado, and Dean}{Mikolov
  et~al\mbox{.}}{2013}]%
        {mikolov2013efficient}
\bibfield{author}{\bibinfo{person}{Tomas Mikolov}, \bibinfo{person}{Kai Chen},
  \bibinfo{person}{Greg Corrado}, {and} \bibinfo{person}{Jeffrey Dean}.}
  \bibinfo{year}{2013}\natexlab{}.
\newblock \showarticletitle{Efficient estimation of word representations in
  vector space}.
\newblock \bibinfo{journal}{\emph{arXiv preprint arXiv:1301.3781}}
  (\bibinfo{year}{2013}).
\newblock


\bibitem[\protect\citeauthoryear{Misra, Wan, and McAuley}{Misra
  et~al\mbox{.}}{2018}]%
        {misra2018decomposing}
\bibfield{author}{\bibinfo{person}{Rishabh Misra}, \bibinfo{person}{Mengting
  Wan}, {and} \bibinfo{person}{Julian McAuley}.}
  \bibinfo{year}{2018}\natexlab{}.
\newblock \showarticletitle{Decomposing fit semantics for product size
  recommendation in metric spaces}. In \bibinfo{booktitle}{\emph{Proceedings of
  the 12th ACM Conference on Recommender Systems}}. \bibinfo{pages}{422--426}.
\newblock


\bibitem[\protect\citeauthoryear{Mohammed~Abdulla, Singh, and
  Borar}{Mohammed~Abdulla et~al\mbox{.}}{2019}]%
        {mohammed2019shop}
\bibfield{author}{\bibinfo{person}{G Mohammed~Abdulla}, \bibinfo{person}{Shreya
  Singh}, {and} \bibinfo{person}{Sumit Borar}.}
  \bibinfo{year}{2019}\natexlab{}.
\newblock \showarticletitle{Shop your Right Size: A System for Recommending
  Sizes for Fashion products}. In \bibinfo{booktitle}{\emph{Companion
  Proceedings of The 2019 World Wide Web Conference}}.
  \bibinfo{pages}{327--334}.
\newblock


\bibitem[\protect\citeauthoryear{Sembium, Rastogi, Saroop, and Merugu}{Sembium
  et~al\mbox{.}}{2017}]%
        {sembium2017recommending}
\bibfield{author}{\bibinfo{person}{Vivek Sembium}, \bibinfo{person}{Rajeev
  Rastogi}, \bibinfo{person}{Atul Saroop}, {and} \bibinfo{person}{Srujana
  Merugu}.} \bibinfo{year}{2017}\natexlab{}.
\newblock \showarticletitle{Recommending product sizes to customers}. In
  \bibinfo{booktitle}{\emph{Proceedings of the Eleventh ACM Conference on
  Recommender Systems}}. \bibinfo{pages}{243--250}.
\newblock


\bibitem[\protect\citeauthoryear{Sembium, Rastogi, Tekumalla, and
  Saroop}{Sembium et~al\mbox{.}}{2018}]%
        {sembium2018bayesian}
\bibfield{author}{\bibinfo{person}{Vivek Sembium}, \bibinfo{person}{Rajeev
  Rastogi}, \bibinfo{person}{Lavanya Tekumalla}, {and} \bibinfo{person}{Atul
  Saroop}.} \bibinfo{year}{2018}\natexlab{}.
\newblock \showarticletitle{Bayesian models for product size recommendations}.
  In \bibinfo{booktitle}{\emph{Proceedings of the 2018 World Wide Web
  Conference}}. \bibinfo{pages}{679--687}.
\newblock


\bibitem[\protect\citeauthoryear{Sheikh, Guigour{\`e}s, Koriagin, Ho, Shirvany,
  Vollgraf, and Bergmann}{Sheikh et~al\mbox{.}}{2019}]%
        {sheikh2019deep}
\bibfield{author}{\bibinfo{person}{Abdul-Saboor Sheikh},
  \bibinfo{person}{Romain Guigour{\`e}s}, \bibinfo{person}{Evgenii Koriagin},
  \bibinfo{person}{Yuen~King Ho}, \bibinfo{person}{Reza Shirvany},
  \bibinfo{person}{Roland Vollgraf}, {and} \bibinfo{person}{Urs Bergmann}.}
  \bibinfo{year}{2019}\natexlab{}.
\newblock \showarticletitle{A deep learning system for predicting size and fit
  in fashion e-commerce}. In \bibinfo{booktitle}{\emph{Proceedings of the 13th
  ACM conference on recommender systems}}. \bibinfo{pages}{110--118}.
\newblock


\bibitem[\protect\citeauthoryear{Statista}{Statista}{2020}]%
        {ecommercereport2020}
\bibfield{author}{\bibinfo{person}{Statista}.} \bibinfo{year}{2020}\natexlab{}.
\newblock \showarticletitle{Fashion eCommerce report 2020. Retrieved Jan 21,
  2021 from}.
\newblock
  \bibinfo{journal}{\emph{https://www.statista.com/study/38340/ecommerce-report-fashion/}}
  (\bibinfo{year}{2020}).
\newblock


\end{thebibliography}

\appendix

\end{document}